\documentstyle[a4,12pt]{article}
\def\beq{\begin{equation}}
\def\eeq{\end{equation}}

\begin{document}

\hfill\hbox{\today}
\vskip 3cm

\begin{center}
{\large{\bf Comment on ``Statistical Mechanics of a Nonlinear
Model for DNA Denaturation''}}

\vspace{2cm}
Su-Long Nyeo and I-Ching Yang
\vspace{5mm}

Department of Physics, National Cheng Kung University\\
Tainan, Taiwan 701, Republic of China
\end{center}

\vspace{2cm}


\vskip 1cm

In a letter [1] Peyrard and Bishop investigated the statistical mechanics of a 
simple lattice model for the denaturation of the DNA double helix.  The 
model consists of two degrees of freedom $u_n$ and $v_n$ corresponding to
the transverse displacements of the bases, i.e., displacements along the 
direction of the hydrogen bonds, which connect the two bases in a pair.  The 
Hamiltonian for the model is given by
\beq
H=\sum_n\frac{1}{2}m({\dot u}_n^2+{\dot v}_n^2)
+\frac{1}{2}k\left[(u_n-u_{n-1})^2+(v_n-v_{n-1})^2\right]+V(u_n-v_n)\,,
\eeq
with the potential between base pairs given by the Morse potential
$$
V(u_n-v_n)=D\left\{\exp[-a(u_n-v_n)]-1\right\}^2\,,
$$
where $D=0.33eV, a=1.8\AA^{-1}$, and the following values for the coupling 
constant $k$ were used: (a) $2\times10^{-3}eV/\AA^2$
(b) $3\times10^{-3}eV/\AA^2$, (c) $4\times10^{-3}eV/\AA^2$.
In terms of the variables 
$x_n=(u_n+v_n)/\sqrt{2}, y_n=(u_n-v_n)/\sqrt{2}$, the
evaluation of the configurational partition function using the transfer 
integral operator method in the thermodynamic limit reduces to solving 
a Schr\"odinger-type equation:
\beq
-\frac{1}{2\beta^2k}\frac{\partial^2\varphi_i(y)}{\partial y^2}
+U_{\rm eff}(y,\beta)\varphi_i(y)
=\epsilon_i\varphi_i(y)\,,\beta=\frac{1}{k_BT}\,,\label{schr}
\eeq
where the temperature-dependent effective potential is given by
\beq
U_{\rm eff}(y,\beta)=D\left\{\exp[-a\sqrt{2}y]-1\right\}^2
+\frac{1}{2\beta}\ln\left(\frac{\beta k}{2\pi}\right)\,.
\eeq
In the thermodynamic limit the ground state dominates with eigen-energy
\beq
\epsilon_0=\frac{1}{2\beta}\ln\left(\frac{\beta k}{2\pi}\right)
+\frac{a}{\beta}\left(\frac{D}{k}\right)^{1/2}
-\frac{a^2}{4\beta^2k}\,,
\eeq
and eigenfunction
\beq
\varphi_0(y)=(\sqrt{2}a)^{1/2}\frac{(2d)^{d-1/2}}{[\Gamma(2d-1)]^{1/2}}
\exp(-de^{-\sqrt{2}ay})\exp[-(d-\frac{1}{2})\sqrt{2}ay]\,.\label{wave}
\eeq
We note that the existence of discrete spectrum is only for values of 
$d\equiv(\beta/a)(kD)^{1/2}$ such that $d>\frac{1}{2}$.  For $d<\frac{1}{2}$, 
equation (\ref{schr}) has only delocalized states.  For the given $k$'s, we 
have the corresponding delocalizing or melting temperatures:
(a) $k=2\times10^{-3}eV/\AA^2\,, T_d=349.868K$,
(b) $k=3\times10^{-3}eV/\AA^2\,, T_d=428.499K$,
(c) $k=4\times10^{-3}eV/\AA^2\,, T_d=494.788K$.
Using the eigenfunction, the authors of [1] evaluated the mean stretching 
$\langle y\rangle=\langle\varphi_0|y|\varphi_0\rangle$ and 
$\langle y^2\rangle=\langle\varphi_0|y^2|\varphi_0\rangle$ beyond the 
delocalizing temperatures, where the $\langle y\rangle$ and 
$\langle y^2\rangle$ in this model are undefined.  The figures given by 
the authors show continuous variations.  In addition, since 
$\langle(y-\langle y\rangle)^2\rangle\ge0$,
we have $\langle y^2\rangle\ge\langle y\rangle^2$.  The figures in [1] do not
reflect this fact.  We have evaluated $\langle y\rangle$ and 
$\langle y^2\rangle$ and show it in {\it FIG. 1}.  
We note from the eigenfunction (\ref{wave}) or from 
{\it FIG. 1} that the hydrogen bonds of between the strands become more 
stretched for larger temperature or smaller $k$.

\begin{figure}
\begin{center}
\setlength{\unitlength}{0.240900pt}
\ifx\plotpoint\undefined\newsavebox{\plotpoint}\fi
\sbox{\plotpoint}{\rule[-0.200pt]{0.400pt}{0.400pt}}%
\begin{picture}(1049,1080)(0,0)
\font\gnuplot=cmr10 at 10pt
\gnuplot
\sbox{\plotpoint}{\rule[-0.200pt]{0.400pt}{0.400pt}}%
\put(220.0,113.0){\rule[-0.200pt]{184.288pt}{0.400pt}}
\put(220.0,113.0){\rule[-0.200pt]{4.818pt}{0.400pt}}
\put(198,113){\makebox(0,0)[r]{0}}
\put(965.0,113.0){\rule[-0.200pt]{4.818pt}{0.400pt}}
\put(220.0,218.0){\rule[-0.200pt]{4.818pt}{0.400pt}}
\put(198,218){\makebox(0,0)[r]{0.5}}
\put(965.0,218.0){\rule[-0.200pt]{4.818pt}{0.400pt}}
\put(220.0,323.0){\rule[-0.200pt]{4.818pt}{0.400pt}}
\put(198,323){\makebox(0,0)[r]{1}}
\put(965.0,323.0){\rule[-0.200pt]{4.818pt}{0.400pt}}
\put(220.0,428.0){\rule[-0.200pt]{4.818pt}{0.400pt}}
\put(198,428){\makebox(0,0)[r]{1.5}}
\put(965.0,428.0){\rule[-0.200pt]{4.818pt}{0.400pt}}
\put(220.0,533.0){\rule[-0.200pt]{4.818pt}{0.400pt}}
\put(198,533){\makebox(0,0)[r]{2}}
\put(965.0,533.0){\rule[-0.200pt]{4.818pt}{0.400pt}}
\put(220.0,637.0){\rule[-0.200pt]{4.818pt}{0.400pt}}
\put(198,637){\makebox(0,0)[r]{2.5}}
\put(965.0,637.0){\rule[-0.200pt]{4.818pt}{0.400pt}}
\put(220.0,742.0){\rule[-0.200pt]{4.818pt}{0.400pt}}
\put(198,742){\makebox(0,0)[r]{3}}
\put(965.0,742.0){\rule[-0.200pt]{4.818pt}{0.400pt}}
\put(220.0,847.0){\rule[-0.200pt]{4.818pt}{0.400pt}}
\put(198,847){\makebox(0,0)[r]{3.5}}
\put(965.0,847.0){\rule[-0.200pt]{4.818pt}{0.400pt}}
\put(220.0,952.0){\rule[-0.200pt]{4.818pt}{0.400pt}}
\put(198,952){\makebox(0,0)[r]{4}}
\put(965.0,952.0){\rule[-0.200pt]{4.818pt}{0.400pt}}
\put(220.0,1057.0){\rule[-0.200pt]{4.818pt}{0.400pt}}
\put(198,1057){\makebox(0,0)[r]{4.5}}
\put(965.0,1057.0){\rule[-0.200pt]{4.818pt}{0.400pt}}
\put(220.0,113.0){\rule[-0.200pt]{0.400pt}{4.818pt}}
\put(220,68){\makebox(0,0){100}}
\put(220.0,1037.0){\rule[-0.200pt]{0.400pt}{4.818pt}}
\put(316.0,113.0){\rule[-0.200pt]{0.400pt}{4.818pt}}
\put(316,68){\makebox(0,0){150}}
\put(316.0,1037.0){\rule[-0.200pt]{0.400pt}{4.818pt}}
\put(411.0,113.0){\rule[-0.200pt]{0.400pt}{4.818pt}}
\put(411,68){\makebox(0,0){200}}
\put(411.0,1037.0){\rule[-0.200pt]{0.400pt}{4.818pt}}
\put(507.0,113.0){\rule[-0.200pt]{0.400pt}{4.818pt}}
\put(507,68){\makebox(0,0){250}}
\put(507.0,1037.0){\rule[-0.200pt]{0.400pt}{4.818pt}}
\put(603.0,113.0){\rule[-0.200pt]{0.400pt}{4.818pt}}
\put(603,68){\makebox(0,0){300}}
\put(603.0,1037.0){\rule[-0.200pt]{0.400pt}{4.818pt}}
\put(698.0,113.0){\rule[-0.200pt]{0.400pt}{4.818pt}}
\put(698,68){\makebox(0,0){350}}
\put(698.0,1037.0){\rule[-0.200pt]{0.400pt}{4.818pt}}
\put(794.0,113.0){\rule[-0.200pt]{0.400pt}{4.818pt}}
\put(794,68){\makebox(0,0){400}}
\put(794.0,1037.0){\rule[-0.200pt]{0.400pt}{4.818pt}}
\put(889.0,113.0){\rule[-0.200pt]{0.400pt}{4.818pt}}
\put(889,68){\makebox(0,0){450}}
\put(889.0,1037.0){\rule[-0.200pt]{0.400pt}{4.818pt}}
\put(985.0,113.0){\rule[-0.200pt]{0.400pt}{4.818pt}}
\put(985,68){\makebox(0,0){500}}
\put(985.0,1037.0){\rule[-0.200pt]{0.400pt}{4.818pt}}
\put(220.0,113.0){\rule[-0.200pt]{184.288pt}{0.400pt}}
\put(985.0,113.0){\rule[-0.200pt]{0.400pt}{227.410pt}}
\put(220.0,1057.0){\rule[-0.200pt]{184.288pt}{0.400pt}}
\put(111,630){\makebox(0,0){\shortstack{$\langle y\rangle$\\$(\AA)$}}}
\put(602,23){\makebox(0,0){$T(K)$}}
\put(564,742){\makebox(0,0)[l]{(a)}}
\put(698,742){\makebox(0,0)[l]{(b)}}
\put(822,742){\makebox(0,0)[l]{(c)}}
\put(220.0,113.0){\rule[-0.200pt]{0.400pt}{227.410pt}}
\put(220,142){\usebox{\plotpoint}}
\multiput(220.00,142.59)(2.475,0.488){13}{\rule{2.000pt}{0.117pt}}
\multiput(220.00,141.17)(33.849,8.000){2}{\rule{1.000pt}{0.400pt}}
\multiput(258.00,150.59)(2.541,0.488){13}{\rule{2.050pt}{0.117pt}}
\multiput(258.00,149.17)(34.745,8.000){2}{\rule{1.025pt}{0.400pt}}
\multiput(297.00,158.59)(2.475,0.488){13}{\rule{2.000pt}{0.117pt}}
\multiput(297.00,157.17)(33.849,8.000){2}{\rule{1.000pt}{0.400pt}}
\multiput(335.00,166.58)(1.955,0.491){17}{\rule{1.620pt}{0.118pt}}
\multiput(335.00,165.17)(34.638,10.000){2}{\rule{0.810pt}{0.400pt}}
\multiput(373.00,176.58)(1.769,0.492){19}{\rule{1.482pt}{0.118pt}}
\multiput(373.00,175.17)(34.924,11.000){2}{\rule{0.741pt}{0.400pt}}
\multiput(411.00,187.58)(1.659,0.492){21}{\rule{1.400pt}{0.119pt}}
\multiput(411.00,186.17)(36.094,12.000){2}{\rule{0.700pt}{0.400pt}}
\multiput(450.00,199.58)(1.488,0.493){23}{\rule{1.269pt}{0.119pt}}
\multiput(450.00,198.17)(35.366,13.000){2}{\rule{0.635pt}{0.400pt}}
\multiput(488.00,212.58)(1.201,0.494){29}{\rule{1.050pt}{0.119pt}}
\multiput(488.00,211.17)(35.821,16.000){2}{\rule{0.525pt}{0.400pt}}
\multiput(526.00,228.58)(1.065,0.495){33}{\rule{0.944pt}{0.119pt}}
\multiput(526.00,227.17)(36.040,18.000){2}{\rule{0.472pt}{0.400pt}}
\multiput(564.00,246.58)(0.934,0.496){39}{\rule{0.843pt}{0.119pt}}
\multiput(564.00,245.17)(37.251,21.000){2}{\rule{0.421pt}{0.400pt}}
\multiput(603.00,267.58)(0.732,0.497){49}{\rule{0.685pt}{0.120pt}}
\multiput(603.00,266.17)(36.579,26.000){2}{\rule{0.342pt}{0.400pt}}
\multiput(641.00,293.58)(0.594,0.497){61}{\rule{0.575pt}{0.120pt}}
\multiput(641.00,292.17)(36.807,32.000){2}{\rule{0.288pt}{0.400pt}}
\multiput(679.58,325.00)(0.498,0.539){73}{\rule{0.120pt}{0.532pt}}
\multiput(678.17,325.00)(38.000,39.897){2}{\rule{0.400pt}{0.266pt}}
\multiput(717.58,366.00)(0.498,0.680){75}{\rule{0.120pt}{0.644pt}}
\multiput(716.17,366.00)(39.000,51.664){2}{\rule{0.400pt}{0.322pt}}
\multiput(756.58,419.00)(0.498,0.924){73}{\rule{0.120pt}{0.837pt}}
\multiput(755.17,419.00)(38.000,68.263){2}{\rule{0.400pt}{0.418pt}}
\multiput(794.58,489.00)(0.498,1.216){73}{\rule{0.120pt}{1.068pt}}
\multiput(793.17,489.00)(38.000,89.782){2}{\rule{0.400pt}{0.534pt}}
\multiput(832.58,581.00)(0.498,1.535){73}{\rule{0.120pt}{1.321pt}}
\multiput(831.17,581.00)(38.000,113.258){2}{\rule{0.400pt}{0.661pt}}
\multiput(870.58,697.00)(0.498,1.741){75}{\rule{0.120pt}{1.485pt}}
\multiput(869.17,697.00)(39.000,131.919){2}{\rule{0.400pt}{0.742pt}}
\multiput(909.58,832.00)(0.498,1.920){73}{\rule{0.120pt}{1.626pt}}
\multiput(908.17,832.00)(38.000,141.625){2}{\rule{0.400pt}{0.813pt}}
\put(220,148){\usebox{\plotpoint}}
\multiput(220.00,148.59)(2.184,0.489){15}{\rule{1.789pt}{0.118pt}}
\multiput(220.00,147.17)(34.287,9.000){2}{\rule{0.894pt}{0.400pt}}
\multiput(258.00,157.58)(2.007,0.491){17}{\rule{1.660pt}{0.118pt}}
\multiput(258.00,156.17)(35.555,10.000){2}{\rule{0.830pt}{0.400pt}}
\multiput(297.00,167.58)(1.769,0.492){19}{\rule{1.482pt}{0.118pt}}
\multiput(297.00,166.17)(34.924,11.000){2}{\rule{0.741pt}{0.400pt}}
\multiput(335.00,178.58)(1.488,0.493){23}{\rule{1.269pt}{0.119pt}}
\multiput(335.00,177.17)(35.366,13.000){2}{\rule{0.635pt}{0.400pt}}
\multiput(373.00,191.58)(1.284,0.494){27}{\rule{1.113pt}{0.119pt}}
\multiput(373.00,190.17)(35.689,15.000){2}{\rule{0.557pt}{0.400pt}}
\multiput(411.00,206.58)(1.159,0.495){31}{\rule{1.018pt}{0.119pt}}
\multiput(411.00,205.17)(36.888,17.000){2}{\rule{0.509pt}{0.400pt}}
\multiput(450.00,223.58)(0.956,0.496){37}{\rule{0.860pt}{0.119pt}}
\multiput(450.00,222.17)(36.215,20.000){2}{\rule{0.430pt}{0.400pt}}
\multiput(488.00,243.58)(0.762,0.497){47}{\rule{0.708pt}{0.120pt}}
\multiput(488.00,242.17)(36.531,25.000){2}{\rule{0.354pt}{0.400pt}}
\multiput(526.00,268.58)(0.634,0.497){57}{\rule{0.607pt}{0.120pt}}
\multiput(526.00,267.17)(36.741,30.000){2}{\rule{0.303pt}{0.400pt}}
\multiput(564.00,298.58)(0.499,0.498){75}{\rule{0.500pt}{0.120pt}}
\multiput(564.00,297.17)(37.962,39.000){2}{\rule{0.250pt}{0.400pt}}
\multiput(603.58,337.00)(0.498,0.698){73}{\rule{0.120pt}{0.658pt}}
\multiput(602.17,337.00)(38.000,51.635){2}{\rule{0.400pt}{0.329pt}}
\multiput(641.58,390.00)(0.498,0.937){73}{\rule{0.120pt}{0.847pt}}
\multiput(640.17,390.00)(38.000,69.241){2}{\rule{0.400pt}{0.424pt}}
\multiput(679.58,461.00)(0.498,1.309){73}{\rule{0.120pt}{1.142pt}}
\multiput(678.17,461.00)(38.000,96.630){2}{\rule{0.400pt}{0.571pt}}
\multiput(717.58,560.00)(0.498,1.663){75}{\rule{0.120pt}{1.423pt}}
\multiput(716.17,560.00)(39.000,126.046){2}{\rule{0.400pt}{0.712pt}}
\multiput(756.58,689.00)(0.498,2.066){73}{\rule{0.120pt}{1.742pt}}
\multiput(755.17,689.00)(38.000,152.384){2}{\rule{0.400pt}{0.871pt}}
\multiput(794.58,845.00)(0.498,2.226){73}{\rule{0.120pt}{1.868pt}}
\multiput(793.17,845.00)(38.000,164.122){2}{\rule{0.400pt}{0.934pt}}
\put(220,158){\usebox{\plotpoint}}
\multiput(220.00,158.58)(1.488,0.493){23}{\rule{1.269pt}{0.119pt}}
\multiput(220.00,157.17)(35.366,13.000){2}{\rule{0.635pt}{0.400pt}}
\multiput(258.00,171.58)(1.318,0.494){27}{\rule{1.140pt}{0.119pt}}
\multiput(258.00,170.17)(36.634,15.000){2}{\rule{0.570pt}{0.400pt}}
\multiput(297.00,186.58)(1.129,0.495){31}{\rule{0.994pt}{0.119pt}}
\multiput(297.00,185.17)(35.937,17.000){2}{\rule{0.497pt}{0.400pt}}
\multiput(335.00,203.58)(0.956,0.496){37}{\rule{0.860pt}{0.119pt}}
\multiput(335.00,202.17)(36.215,20.000){2}{\rule{0.430pt}{0.400pt}}
\multiput(373.00,223.58)(0.732,0.497){49}{\rule{0.685pt}{0.120pt}}
\multiput(373.00,222.17)(36.579,26.000){2}{\rule{0.342pt}{0.400pt}}
\multiput(411.00,249.58)(0.609,0.497){61}{\rule{0.588pt}{0.120pt}}
\multiput(411.00,248.17)(37.781,32.000){2}{\rule{0.294pt}{0.400pt}}
\multiput(450.58,281.00)(0.498,0.565){73}{\rule{0.120pt}{0.553pt}}
\multiput(449.17,281.00)(38.000,41.853){2}{\rule{0.400pt}{0.276pt}}
\multiput(488.58,324.00)(0.498,0.804){73}{\rule{0.120pt}{0.742pt}}
\multiput(487.17,324.00)(38.000,59.460){2}{\rule{0.400pt}{0.371pt}}
\multiput(526.58,385.00)(0.498,1.176){73}{\rule{0.120pt}{1.037pt}}
\multiput(525.17,385.00)(38.000,86.848){2}{\rule{0.400pt}{0.518pt}}
\multiput(564.58,474.00)(0.498,1.676){75}{\rule{0.120pt}{1.433pt}}
\multiput(563.17,474.00)(39.000,127.025){2}{\rule{0.400pt}{0.717pt}}
\multiput(603.58,604.00)(0.498,2.332){73}{\rule{0.120pt}{1.953pt}}
\multiput(602.17,604.00)(38.000,171.947){2}{\rule{0.400pt}{0.976pt}}
\multiput(641.58,780.00)(0.498,2.690){73}{\rule{0.120pt}{2.237pt}}
\multiput(640.17,780.00)(38.000,198.357){2}{\rule{0.400pt}{1.118pt}}
\end{picture}
\setlength{\unitlength}{0.240900pt}
\ifx\plotpoint\undefined\newsavebox{\plotpoint}\fi
\sbox{\plotpoint}{\rule[-0.200pt]{0.400pt}{0.400pt}}%
\begin{picture}(1049,1080)(0,0)
\font\gnuplot=cmr10 at 10pt
\gnuplot
\sbox{\plotpoint}{\rule[-0.200pt]{0.400pt}{0.400pt}}%
\put(220.0,113.0){\rule[-0.200pt]{184.288pt}{0.400pt}}
\put(220.0,113.0){\rule[-0.200pt]{4.818pt}{0.400pt}}
\put(198,113){\makebox(0,0)[r]{0}}
\put(965.0,113.0){\rule[-0.200pt]{4.818pt}{0.400pt}}
\put(220.0,302.0){\rule[-0.200pt]{4.818pt}{0.400pt}}
\put(198,302){\makebox(0,0)[r]{5}}
\put(965.0,302.0){\rule[-0.200pt]{4.818pt}{0.400pt}}
\put(220.0,491.0){\rule[-0.200pt]{4.818pt}{0.400pt}}
\put(198,491){\makebox(0,0)[r]{10}}
\put(965.0,491.0){\rule[-0.200pt]{4.818pt}{0.400pt}}
\put(220.0,679.0){\rule[-0.200pt]{4.818pt}{0.400pt}}
\put(198,679){\makebox(0,0)[r]{15}}
\put(965.0,679.0){\rule[-0.200pt]{4.818pt}{0.400pt}}
\put(220.0,868.0){\rule[-0.200pt]{4.818pt}{0.400pt}}
\put(198,868){\makebox(0,0)[r]{20}}
\put(965.0,868.0){\rule[-0.200pt]{4.818pt}{0.400pt}}
\put(220.0,1057.0){\rule[-0.200pt]{4.818pt}{0.400pt}}
\put(198,1057){\makebox(0,0)[r]{25}}
\put(965.0,1057.0){\rule[-0.200pt]{4.818pt}{0.400pt}}
\put(220.0,113.0){\rule[-0.200pt]{0.400pt}{4.818pt}}
\put(220,68){\makebox(0,0){100}}
\put(220.0,1037.0){\rule[-0.200pt]{0.400pt}{4.818pt}}
\put(316.0,113.0){\rule[-0.200pt]{0.400pt}{4.818pt}}
\put(316,68){\makebox(0,0){150}}
\put(316.0,1037.0){\rule[-0.200pt]{0.400pt}{4.818pt}}
\put(411.0,113.0){\rule[-0.200pt]{0.400pt}{4.818pt}}
\put(411,68){\makebox(0,0){200}}
\put(411.0,1037.0){\rule[-0.200pt]{0.400pt}{4.818pt}}
\put(507.0,113.0){\rule[-0.200pt]{0.400pt}{4.818pt}}
\put(507,68){\makebox(0,0){250}}
\put(507.0,1037.0){\rule[-0.200pt]{0.400pt}{4.818pt}}
\put(603.0,113.0){\rule[-0.200pt]{0.400pt}{4.818pt}}
\put(603,68){\makebox(0,0){300}}
\put(603.0,1037.0){\rule[-0.200pt]{0.400pt}{4.818pt}}
\put(698.0,113.0){\rule[-0.200pt]{0.400pt}{4.818pt}}
\put(698,68){\makebox(0,0){350}}
\put(698.0,1037.0){\rule[-0.200pt]{0.400pt}{4.818pt}}
\put(794.0,113.0){\rule[-0.200pt]{0.400pt}{4.818pt}}
\put(794,68){\makebox(0,0){400}}
\put(794.0,1037.0){\rule[-0.200pt]{0.400pt}{4.818pt}}
\put(889.0,113.0){\rule[-0.200pt]{0.400pt}{4.818pt}}
\put(889,68){\makebox(0,0){450}}
\put(889.0,1037.0){\rule[-0.200pt]{0.400pt}{4.818pt}}
\put(985.0,113.0){\rule[-0.200pt]{0.400pt}{4.818pt}}
\put(985,68){\makebox(0,0){500}}
\put(985.0,1037.0){\rule[-0.200pt]{0.400pt}{4.818pt}}
\put(220.0,113.0){\rule[-0.200pt]{184.288pt}{0.400pt}}
\put(985.0,113.0){\rule[-0.200pt]{0.400pt}{227.410pt}}
\put(220.0,1057.0){\rule[-0.200pt]{184.288pt}{0.400pt}}
\put(111,630){\makebox(0,0){\shortstack{$\langle y^2\rangle$\\$(\AA^2)$}}}
\put(602,23){\makebox(0,0){$T(K)$}}
\put(564,679){\makebox(0,0)[l]{(a)}}
\put(698,679){\makebox(0,0)[l]{(b)}}
\put(822,679){\makebox(0,0)[l]{(c)}}
\put(220.0,113.0){\rule[-0.200pt]{0.400pt}{227.410pt}}
\put(220,115){\usebox{\plotpoint}}
\put(220,114.67){\rule{9.154pt}{0.400pt}}
\multiput(220.00,114.17)(19.000,1.000){2}{\rule{4.577pt}{0.400pt}}
\put(258,116.17){\rule{7.900pt}{0.400pt}}
\multiput(258.00,115.17)(22.603,2.000){2}{\rule{3.950pt}{0.400pt}}
\put(297,117.67){\rule{9.154pt}{0.400pt}}
\multiput(297.00,117.17)(19.000,1.000){2}{\rule{4.577pt}{0.400pt}}
\put(335,119.17){\rule{7.700pt}{0.400pt}}
\multiput(335.00,118.17)(22.018,2.000){2}{\rule{3.850pt}{0.400pt}}
\put(373,121.17){\rule{7.700pt}{0.400pt}}
\multiput(373.00,120.17)(22.018,2.000){2}{\rule{3.850pt}{0.400pt}}
\multiput(411.00,123.61)(8.500,0.447){3}{\rule{5.300pt}{0.108pt}}
\multiput(411.00,122.17)(28.000,3.000){2}{\rule{2.650pt}{0.400pt}}
\multiput(450.00,126.60)(5.453,0.468){5}{\rule{3.900pt}{0.113pt}}
\multiput(450.00,125.17)(29.905,4.000){2}{\rule{1.950pt}{0.400pt}}
\multiput(488.00,130.59)(4.161,0.477){7}{\rule{3.140pt}{0.115pt}}
\multiput(488.00,129.17)(31.483,5.000){2}{\rule{1.570pt}{0.400pt}}
\multiput(526.00,135.59)(2.857,0.485){11}{\rule{2.271pt}{0.117pt}}
\multiput(526.00,134.17)(33.286,7.000){2}{\rule{1.136pt}{0.400pt}}
\multiput(564.00,142.58)(2.007,0.491){17}{\rule{1.660pt}{0.118pt}}
\multiput(564.00,141.17)(35.555,10.000){2}{\rule{0.830pt}{0.400pt}}
\multiput(603.00,152.58)(1.378,0.494){25}{\rule{1.186pt}{0.119pt}}
\multiput(603.00,151.17)(35.539,14.000){2}{\rule{0.593pt}{0.400pt}}
\multiput(641.00,166.58)(1.008,0.495){35}{\rule{0.900pt}{0.119pt}}
\multiput(641.00,165.17)(36.132,19.000){2}{\rule{0.450pt}{0.400pt}}
\multiput(679.00,185.58)(0.634,0.497){57}{\rule{0.607pt}{0.120pt}}
\multiput(679.00,184.17)(36.741,30.000){2}{\rule{0.303pt}{0.400pt}}
\multiput(717.58,215.00)(0.498,0.590){75}{\rule{0.120pt}{0.572pt}}
\multiput(716.17,215.00)(39.000,44.813){2}{\rule{0.400pt}{0.286pt}}
\multiput(756.58,261.00)(0.498,0.924){73}{\rule{0.120pt}{0.837pt}}
\multiput(755.17,261.00)(38.000,68.263){2}{\rule{0.400pt}{0.418pt}}
\multiput(794.58,331.00)(0.498,1.402){73}{\rule{0.120pt}{1.216pt}}
\multiput(793.17,331.00)(38.000,103.477){2}{\rule{0.400pt}{0.608pt}}
\multiput(832.58,437.00)(0.498,1.973){73}{\rule{0.120pt}{1.668pt}}
\multiput(831.17,437.00)(38.000,145.537){2}{\rule{0.400pt}{0.834pt}}
\multiput(870.58,586.00)(0.498,2.466){75}{\rule{0.120pt}{2.059pt}}
\multiput(869.17,586.00)(39.000,186.726){2}{\rule{0.400pt}{1.029pt}}
\multiput(909.58,777.00)(0.498,2.930){73}{\rule{0.120pt}{2.426pt}}
\multiput(908.17,777.00)(38.000,215.964){2}{\rule{0.400pt}{1.213pt}}
\put(220,116){\usebox{\plotpoint}}
\put(220,115.67){\rule{9.154pt}{0.400pt}}
\multiput(220.00,115.17)(19.000,1.000){2}{\rule{4.577pt}{0.400pt}}
\put(258,117.17){\rule{7.900pt}{0.400pt}}
\multiput(258.00,116.17)(22.603,2.000){2}{\rule{3.950pt}{0.400pt}}
\put(297,119.17){\rule{7.700pt}{0.400pt}}
\multiput(297.00,118.17)(22.018,2.000){2}{\rule{3.850pt}{0.400pt}}
\multiput(335.00,121.61)(8.276,0.447){3}{\rule{5.167pt}{0.108pt}}
\multiput(335.00,120.17)(27.276,3.000){2}{\rule{2.583pt}{0.400pt}}
\multiput(373.00,124.60)(5.453,0.468){5}{\rule{3.900pt}{0.113pt}}
\multiput(373.00,123.17)(29.905,4.000){2}{\rule{1.950pt}{0.400pt}}
\multiput(411.00,128.59)(3.474,0.482){9}{\rule{2.700pt}{0.116pt}}
\multiput(411.00,127.17)(33.396,6.000){2}{\rule{1.350pt}{0.400pt}}
\multiput(450.00,134.59)(2.857,0.485){11}{\rule{2.271pt}{0.117pt}}
\multiput(450.00,133.17)(33.286,7.000){2}{\rule{1.136pt}{0.400pt}}
\multiput(488.00,141.58)(1.769,0.492){19}{\rule{1.482pt}{0.118pt}}
\multiput(488.00,140.17)(34.924,11.000){2}{\rule{0.741pt}{0.400pt}}
\multiput(526.00,152.58)(1.201,0.494){29}{\rule{1.050pt}{0.119pt}}
\multiput(526.00,151.17)(35.821,16.000){2}{\rule{0.525pt}{0.400pt}}
\multiput(564.00,168.58)(0.752,0.497){49}{\rule{0.700pt}{0.120pt}}
\multiput(564.00,167.17)(37.547,26.000){2}{\rule{0.350pt}{0.400pt}}
\multiput(603.58,194.00)(0.498,0.526){73}{\rule{0.120pt}{0.521pt}}
\multiput(602.17,194.00)(38.000,38.919){2}{\rule{0.400pt}{0.261pt}}
\multiput(641.58,234.00)(0.498,0.897){73}{\rule{0.120pt}{0.816pt}}
\multiput(640.17,234.00)(38.000,66.307){2}{\rule{0.400pt}{0.408pt}}
\multiput(679.58,302.00)(0.498,1.442){73}{\rule{0.120pt}{1.247pt}}
\multiput(678.17,302.00)(38.000,106.411){2}{\rule{0.400pt}{0.624pt}}
\multiput(717.58,411.00)(0.498,2.116){75}{\rule{0.120pt}{1.782pt}}
\multiput(716.17,411.00)(39.000,160.301){2}{\rule{0.400pt}{0.891pt}}
\multiput(756.58,575.00)(0.498,2.943){73}{\rule{0.120pt}{2.437pt}}
\multiput(755.17,575.00)(38.000,216.942){2}{\rule{0.400pt}{1.218pt}}
\multiput(794.58,797.00)(0.498,3.434){73}{\rule{0.120pt}{2.826pt}}
\multiput(793.17,797.00)(38.000,253.134){2}{\rule{0.400pt}{1.413pt}}
\put(220,118){\usebox{\plotpoint}}
\put(220,118.17){\rule{7.700pt}{0.400pt}}
\multiput(220.00,117.17)(22.018,2.000){2}{\rule{3.850pt}{0.400pt}}
\multiput(258.00,120.61)(8.500,0.447){3}{\rule{5.300pt}{0.108pt}}
\multiput(258.00,119.17)(28.000,3.000){2}{\rule{2.650pt}{0.400pt}}
\multiput(297.00,123.60)(5.453,0.468){5}{\rule{3.900pt}{0.113pt}}
\multiput(297.00,122.17)(29.905,4.000){2}{\rule{1.950pt}{0.400pt}}
\multiput(335.00,127.59)(2.857,0.485){11}{\rule{2.271pt}{0.117pt}}
\multiput(335.00,126.17)(33.286,7.000){2}{\rule{1.136pt}{0.400pt}}
\multiput(373.00,134.58)(1.955,0.491){17}{\rule{1.620pt}{0.118pt}}
\multiput(373.00,133.17)(34.638,10.000){2}{\rule{0.810pt}{0.400pt}}
\multiput(411.00,144.58)(1.318,0.494){27}{\rule{1.140pt}{0.119pt}}
\multiput(411.00,143.17)(36.634,15.000){2}{\rule{0.570pt}{0.400pt}}
\multiput(450.00,159.58)(0.732,0.497){49}{\rule{0.685pt}{0.120pt}}
\multiput(450.00,158.17)(36.579,26.000){2}{\rule{0.342pt}{0.400pt}}
\multiput(488.58,185.00)(0.498,0.592){73}{\rule{0.120pt}{0.574pt}}
\multiput(487.17,185.00)(38.000,43.809){2}{\rule{0.400pt}{0.287pt}}
\multiput(526.58,230.00)(0.498,1.123){73}{\rule{0.120pt}{0.995pt}}
\multiput(525.17,230.00)(38.000,82.935){2}{\rule{0.400pt}{0.497pt}}
\multiput(564.58,315.00)(0.498,1.935){75}{\rule{0.120pt}{1.638pt}}
\multiput(563.17,315.00)(39.000,146.599){2}{\rule{0.400pt}{0.819pt}}
\multiput(603.58,465.00)(0.498,3.129){73}{\rule{0.120pt}{2.584pt}}
\multiput(602.17,465.00)(38.000,230.636){2}{\rule{0.400pt}{1.292pt}}
\multiput(641.58,701.00)(0.498,4.072){73}{\rule{0.120pt}{3.332pt}}
\multiput(640.17,701.00)(38.000,300.085){2}{\rule{0.400pt}{1.666pt}}
\end{picture}
{\it FIG. 1:  Variation of $\langle y\rangle$ and of $\langle y^2\rangle$ 
as a function of temperature for three values of the coupling constant $k$: 
(a) $k=2.0\times10^{-3} eV/\AA^2$, (b)$k=3.0\times10^{-3}eV/\AA^2$,(c)
$k=4.0\times10^{-3}eV/\AA^2$.}
\end{center}
\end{figure}

\section*{Acknowledgments}

\noindent This research was supported by the National Science Council of the
Republic of China under Contract Nos. NSC 87-2112-M-006-002
and NSC 88-2112-M-006-003.

\end{document}